\documentclass[]{aa}
\usepackage{graphicx}
\usepackage{txfonts}
\usepackage[colorlinks]{hyperref}
\hypersetup{colorlinks=true,linkcolor=blue,citecolor=blue,filecolor=blue,urlcolor=blue,}
\usepackage{amsmath}
\usepackage{algorithm}
\usepackage{amsfonts}
\usepackage{mathrsfs}
\usepackage{comment}
\usepackage{bm}
\usepackage{rotating}
\usepackage{color}
\usepackage{graphicx}
\usepackage{subfigure}
\usepackage{savesym}
\usepackage[flushleft]{threeparttable}
\savesymbol{tablenum}
\restoresymbol{SIX}{tablenum}

\definecolor{amethyst}{rgb}{0.8, 0.0, 0.0}

\usepackage{lineno}

\def \a3 {A$^3$COSMOS }

\def \md {$M_{\rm D}$ }

\begin{document}

   \title{A$^3$COSMOS: the dust mass function and dust mass density at $0.5<z<6$}

    \author{A. Traina \inst{1,2}, B. Magnelli \inst{3}, C. Gruppioni \inst{1}, I. Delvecchio \inst{4}, M. Parente \inst{5,6}, F. Calura \inst{1}, L. Bisigello \inst{7,8}, A. Feltre \inst{9},  F. Pozzi \inst{1,2}, L. Vallini \inst{1}
          }

   \institute{Istituto Nazionale di Astrofisica (INAF) - Osservatorio di Astrofisica e Scienza dello Spazio (OAS), via Gobetti 101, I-40129 Bologna, Italy
         \and
             Dipartimento di Fisica e Astronomia (DIFA), Universit\`a di Bologna, via Gobetti 93/2, I-40129 Bologna, Italy
         \and 
             Université Paris-Saclay, Université Paris  Cité, CEA, CNRS, AIM, F-91191 Gif-sur-Yvette, France
         \and
             INAF, Osservatorio Astronomico di Brera 28, 20121, Milano, Italy and Via Bianchi 46, 23807 Merate, Italy
         \and
             SISSA, Via Bonomea 265, 34136 Trieste, Italy
         \and
             INAF, Osservatorio Astronomico di Trieste, via Tiepolo 11, I-34131, Trieste, Italy
         \and
             Dipartimento di Fisica e Astronomia "G. Galilei", Universit\`a di Padova, Via Marzolo 8, 35131 Padova, Italy
         \and 
             INAF, Istituto di Radioastronomia, Via Piero Gobetti 101, 40129 Bologna, Italy
         \and 
             INAF - Osservatorio Astrofisico di Arcetri, Largo E. Fermi 5, 50125, Firenze, Italy
             }

   \date{Received 14 June 2024; accepted 11 July 2024}

 
  \abstract
   {Although dust in galaxies represents only a few percent of the total baryonic mass, it plays a crucial role in the physical processes occurring in galaxies. Studying the dust content of galaxies, particularly at high$-z$, is therefore crucial to understand the link between dust production, obscured star formation and the build-up of galaxy stellar mass.}
   {To study the dust properties (mass and temperature) of the largest Atacama Large Millimeter/submillimeter Array (ALMA)-selected sample of star-forming galaxies available from the archive (A$^3$COSMOS) and derive the dust mass function and dust mass density of galaxies from $z=0.5\,-\,6$.}
 {We performed spectral energy distribution (SED) fitting with the \texttt{CIGALE} code to constrain the dust mass and temperature of the A$^3$COSMOS galaxy sample, thanks to the UV-to-near-infrared photometric coverage of each galaxies combined with the ALMA (and {\it Herschel} when available) coverage of the Rayleigh-Jeans tail of their dust-continuum emission. We then computed and fitted the dust mass function by combining the A$^3$COSMOS and state-of-the-art {\it Herschel} samples, in order to obtain the best estimate of the integrated dust mass density up to $z \sim 6$.}
    {Galaxies in \a3 have dust masses between $\sim 10^8$ and $\sim 10^{9.5}$ M$_{\odot}$. From the SED fitting, we were also able to derive a dust temperature, finding that the distribution of the dust temperature peaks at $\sim 30-35$K. The dust mass function at $z=0.5\,-\,6$ evolves with an increase of $M^*$ and decrease of the number density ($\Phi ^*$) and is in good agreement with literature estimates. The dust mass density shows a smooth decrease in its evolution from $z \sim 0.5$ to $z \sim 6$, which is steeper than what is found by models at $z \gtrsim 2$.}  
   {}


   \titlerunning{The DMF and DMD at $z \sim 0.5-6$ from the A$^3$COSMOS survey}
   \authorrunning{A. Traina et al.}
   \maketitle   
%

\section{Introduction}\label{sec:intro}

Recent studies have reveal a larger value of dust-obscured star formation rate density (SFRD) even at large redshift \citep[e.g.,][]{khusanova2021sfrd,gruppioni2020alpine,Traina2024sfrd}.
To understand the origin of such large dust-obscured SFRD a possibility is to study the dust mass content of the Universe at different cosmic times. Such a study is not only crucial for our understanding of dust-obscured SFRD, but also because dust plays a major role in most astrophysical processes in the evolution of galaxies and active galactic nuclei (AGN) \citep{mckee2007sf,cazaux2002ism,hopkins2012stellarfeedback}, and why and when dust emerges is still debated. Deriving the dust mass function (DMF) and the evolution with redshift of the dust mass density (DMD) offers the opportunity to address this issue via a statistical approach.
\par In the past years, a number of works have been devoted to the estimation of the dust mass in galaxies and how it scales with other galactic properties in local and distant galaxies \citep{calura2017dust_prop,Herrera-Camus2018dustprop,pastrav2020dustprop,casasola2022dustpedia}, mostly using the {\it Herschel} observatory \citep[]{dunne2011dmd, pozzi2020dmd, eales2024dmd_herschel}. However, while these studies were able to trace the DMD from $z \sim 2-3$ to the present days, they did not probe its evolution at higher redshifts, because of the limitation of the far-infrared (FIR) in tracing the Rayleigh-Jeans (hereafter R-J) regime of the dust emission (needed for deriving \md). Sampling the dust emission at longer wavelengths such as the sub-millimeter (sub-mm) and millimeter (mm) bands is then crucial to explore the high-$z$ dust content of the Universe. 
Different studies, using sub-mm/mm facilities (e.g., IRAM, ALMA) have investigated the dust mass density at higher redshifts \citep[]{magnelli2019dmd,magnelli2020dmd,pozzi2021dmd} and their estimates are pointing toward different contents of dust at high-$z$. Moreover, simulations \citep[][]{popping2017dmd_sim,aoyama2018dmd_sim,li2019dmd_sim,vijayan2019dmd_sim} fail to accurately reproduce the observed evolution of the DMD with redshift, and in particular very few of them are able to reproduce the observed drop at $z<1$ \citep[]{gioannini2017dmd_sim,parente2023dmdm_sim}. In this context, deep sub-mm surveys of large galaxy samples and covering a wide redshift range  are critical in order to shed light on the evolution of galaxy's dust content across cosmic time. 
\par Although ALMA is characterized by a small field of view, which makes wide-area surveys too observationally demanding, the recent A$^3$COSMOS survey \citep{liu2019a32,liu2019a31,adscheid2024a3cosmos} circumvents this limitation by collecting and homogeneously analysing archival ALMA images of dusty star-forming galaxies. This unique survey can also be exploited for characterizing the dust mass content and evolution. Even though the \a3 is by construction an inhomogeneous collection of ALMA archival observations, significant efforts have been invested to statistically turn it from a pointed to a “blind-like” survey (reducing possible bias sources). Such effort has enable a variety of relevant statistical quantities to be derived \citep[i.e., number counts, luminosity function, SFRD;][]{adscheid2024a3cosmos,Traina2024sfrd}. Continuing exploiting the A$^3$COSMOS database, in this work we characterize their dust properties and estimate the DMF and DMD over a wide ($0.5-6$) redshift range.
\par The paper is organized as follows. In Section \ref{sec:sample} we present the \a3 sample; in Section \ref{sec:mdust} we describe the methods used to derive the \md and discuss their differences; in Section \ref{sec:DMFD} we present the results on the DMF and DMD, which we compare to predictions from semi-analytical models and simulations in Section \ref{discussion}. Throughout the present work, we assume a \cite{chabrier2003imf} stellar
initial mass function (IMF) and adopt a $\Lambda$CDM cosmology with $H_{0} = 70$ $\rm km$ $\rm s^{-1}$ $\rm Mpc^{-1}$, $\Omega_{\rm m} = 0.3$, and $\Omega_{\Lambda} = 0.7$.

%
%
\section{Data sample: the A3COSMOS database}\label{sec:sample} 

The \a3 survey, described by \cite{liu2019a31} and recently updated by \cite{adscheid2024a3cosmos}, consists in the collection of all the archival ALMA observations within the COSMOS field. In this paper, we used the most recent version of the \a3 combined with the COSMOS2020 photometry \citep{adscheid2024a3cosmos}, which was already adopted in \cite{Traina2024sfrd}. Due to the inhomogeneous nature of this database, with each ALMA pointing having its observing frequency, sensitivity and particular targeted source, a reduction of all the possible biases affecting the sample is needed. In this work, we used the final sample obtained by \cite{Traina2024sfrd} \citep[see also][]{adscheid2024a3cosmos}, in which a “blinding” process was applied, in order to use this database as a blind survey for statistical purposes. Firstly, to account for possible positional errors between the ALMA target and sub-mm priors from other catalogues, all the pointings without a source in the inner 1” were removed, as well as the area of the inner circle. Clustered galaxies, with a redshift similar to that of the target, were also removed. Finally, specific conversions to account for different wavelenghts and sensitivities were applied, during the calculation of the areal coverage for the so-built A$^3$COSMOS survey. The final sample consists of 189 galaxies which are considered serendipitously detected in the ALMA pointings. 
\par The main integrated galaxy properties (i.e., stellar mass, dust luminosity, star formation rate) were inferred by \cite{Traina2024sfrd}, finding a massive ($M_{\star} \sim 10^{10}-10^{12}$ M$_{\odot}$), infrared luminous ($L_{\rm IR,8-1000} \sim 10^{11}-10^{13}$ L$_{\odot}$) and highly star-forming (SFR $\sim 10-1000$ M$_{\odot}$yr$^{-1}$) population. Also, from the SED fitting performed in \cite{Traina2024sfrd} (see Fig. 4 in the paper), we find that $\sim 30-40\%$ of these galaxies are likely hosting an AGN, with the bulk of them having a low AGN contribution to the total SED emission (i.e., fraction of the AGN emission to the total IR emission in the 5-40 $\mu$m range, $f_{\rm AGN}$, with a mean value of $\sim 35\%$). This sample of star forming galaxies having a sub-mm/mm detection with ALMA, is particularly well suited for the characterization of the dust properties of SFGs.  

%
%

\section{Dust mass estimation}\label{sec:mdust} 

The IR emission in galaxies originates from dust heated either by UV emission from recently formed stars or from the AGN. The dust thermalizes with the radiation and re-emits at longer wavelengths, in the IR bands, with a spectrum that can reliably described by a grey body emission \citep[i.e., less efficient than a black body,][]{bianchi2013dust}, whose shape is linked to the different dust phases in a galaxy. The warm dust component, close to the heating source (stars or AGN) emits large quantities of luminosity per unit of mass and dominates the FIR emission (where dust emission peaks). The diffuse dust component in the galaxy, which dominates in mass, is instead heated by a weaker radiation field, thus having lower temperatures. This implies a much lower luminosity per unit mass and an emission peaked at longer wavelengths (sub-mm/mm). The composite contribution of the different dust phases produces the observed peak of the dust emission ($\lambda_{\rm peak}$). At $\lambda > \lambda_{\rm peak}$, the grey body emission is in the R-J regime (where cold dust peaks) and can be used to trace the global dust mass in the galaxy. In particular, the R-J emission traces the bulk of the dust reservoir in the galaxy, which is at the “mass-weighted” temperature, in contrast to the “luminosity-weighted” temperature, that would be associated to strongly heated dust, emitting in the FIR, but not representative of the total dust mass \citep[see][for a detailed discussion on the different dust temperature definitions]{liang2019tdust}. In this work, we compute the \md relying on the SED fitting or directly from the observed ALMA flux. In the following Sections \ref{subsec:md_sed} and \ref{subsec:md_flux}, we describe these two possible approaches to derive $M_{\rm D}$.

\subsection{$M_{\rm D}$ from SED fitting} \label{subsec:md_sed}
\begin{figure}[]
\centering
{\includegraphics[width=.5\textwidth]{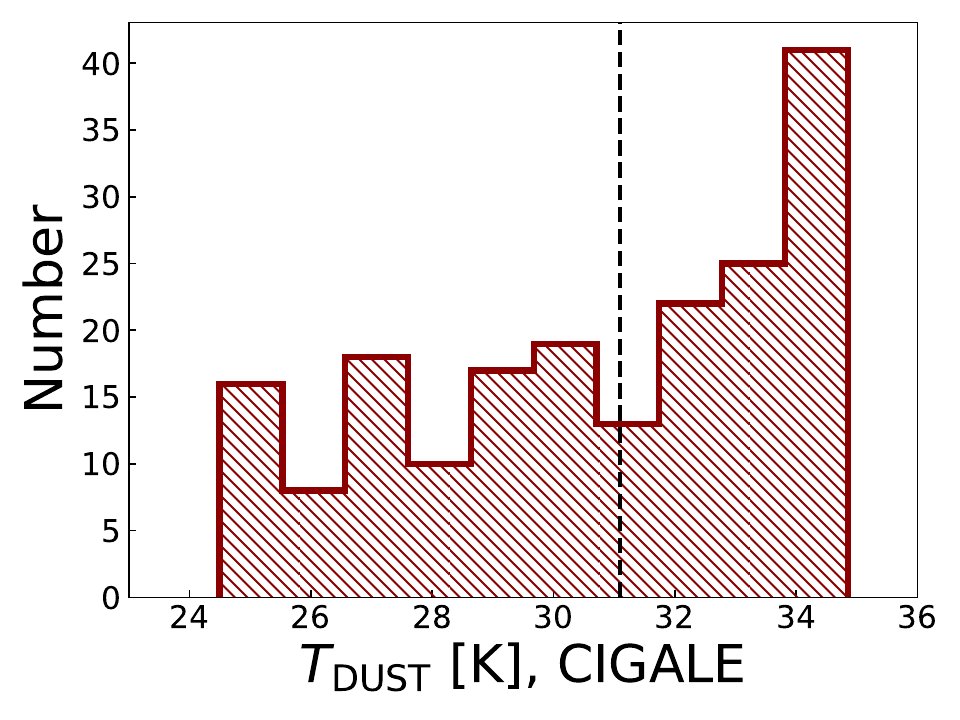}}
\caption{\small{Dust temperature computed using the output radiation field given by the \texttt{CIGALE} SED fitting analysis (see Eq. \ref{eq:tdust}). The vertical dashed line indicates the median value for the sample.}}
\label{fig:tdust_cigale}
\end{figure}

The first method to infer \md of a galaxy is to perform a complete SED fitting of the FIR/mm photometry, that is tracing dust emission. In this paper, we use the \textsc{python} “Code Investigating GALaxy Emission" \citep[\texttt{CIGALE};][]{boquien2019cigale} SED-fitting tool. \texttt{CIGALE} is based on the energy balance between the UV and optical emission by stars and the re-emission in the IR and mm by the dust and it allows one to choose among different individual templates for each emission component (e.g., stellar optical/UV emission, cold dust emission, AGN emission) across a broad parameter space. For the dust emission component, we use the \texttt{DL14} \citep{draine2014dustem} module. Consistently with the dust attenuation component used, \texttt{DL14} assumes the presence of both the diffuse interstellar medium (ISM), heated by the general stellar population in the galaxy, and birth clouds (BCs), heated by newly formed massive stars. The main input parameters are the PAH mass fraction and the minimum radiation field, along with the slope of the dust emission ($\alpha$) and the illuminated fraction ($\gamma$), see Table A.1 in \cite{Traina2024sfrd} for the input parameters. The dust mass of the cold component is computed by dividing the dust luminosity (i.e., the luminosity arising from the dust heated by stars) by the dust emissivity, and it is returned as output from the SED-fitting. The radiation field ($U_{\rm min}$, returned from \texttt{CIGALE}) that is responsible for heating the dust can be linked to the mass-weighted dust temperature as follows \citep{aniano2012tdust}:
\begin{equation} \label{eq:tdust}
    T_{\rm D, CIGALE} = 18 \cdot U_{\rm min}^{1/6} {\rm [K]}.
\end{equation}
Figure \ref{fig:tdust_cigale} shows the $T_{\rm D, CIGALE}$ distribution, which follows a smoothly rising trend, with a large number of sources having $T_{\rm D, CIGALE} \sim 34$K. The median value is $\sim 31$K, in agreement to what is found by \cite{pozzi2020dmd} from a sample of $\sim 6000$ galaxies observed with {\it Herschel} at $z \sim 0-2.5$.

\subsection{$M_{\rm D}$ from R-J flux density} \label{subsec:md_flux}

An alternative method to derive \md consists in using the flux at a certain wavelength in the R-J part of the SED. Following the optically thin approximation, \md can be computed as follows:
\begin{equation} \label{eq:mdust}
 \centering
    M_{\rm D} = \frac{5.03 \times 10^{-31} \cdot S_{\nu_{\rm obs}} \cdot D_{\rm L}^2}{(1+z)^4 \cdot B_{\nu_{\rm obs}}(T_{\rm obs}) \cdot \kappa_{\nu_o}} \cdot \left( \frac{\nu_o}{\nu_{\rm rest}}  \right)^{\beta} ;
\end{equation}
where $ B_{\nu_{\rm obs}}(T_{\rm obs})$ is the black-body Planck function computed at the observed frame, mass-weighted, temperature ($T_{\rm obs} = T_{\rm D} / (1+z)$);  $S_{\nu_{\rm obs}}$ is the flux density at the observed frequency $\nu_{\rm obs}$, with $\nu_{\rm rest} = (1+z) \nu_{\rm obs}$; $\kappa_{\nu_o}$ is the photon cross-section to mass ratio of dust at the rest-frame $\nu_0$ and $\beta$ is the dust emissivity spectral index \citep[see e.g.,][]{magnelli2020dmd}{}{}. he approximation of being in the optically-thin regime has already been assumed in several works in the literature \citep[see e.g.,][]{magnelli2020dmd,pozzi2021dmd} and has been proved to be valid when the bulk of the galaxy population has a cold dust component (as in our case) and when the observed flux is in the R-J regime \citep{scoville2014dust,scoville2016dust_mass,scoville2017alma}. In our analysis, we adopted a value of $\kappa = 0.0469$ m$^2$kg$^{-1}$, derived for a wavelength of $850 \mu$m \citep[][]{draine2014dustem} to be consistent with the \texttt{CIGALE} estimates. For the spectral index of the dust emissivity, $\beta$, typical values are between 1.5 and 2.0 \citep[with a good agreement between empirical measurements and theoretical predictions][]{dunne2001dust,clements2010dust,draine2011book}. Here we used $\beta = 1.8$, as suggested value by \cite{scoville2014dust} based on the findings of the \cite{planck_collab2011}.
\par
\begin{figure}[]
\centering
{\includegraphics[width=.5\textwidth]{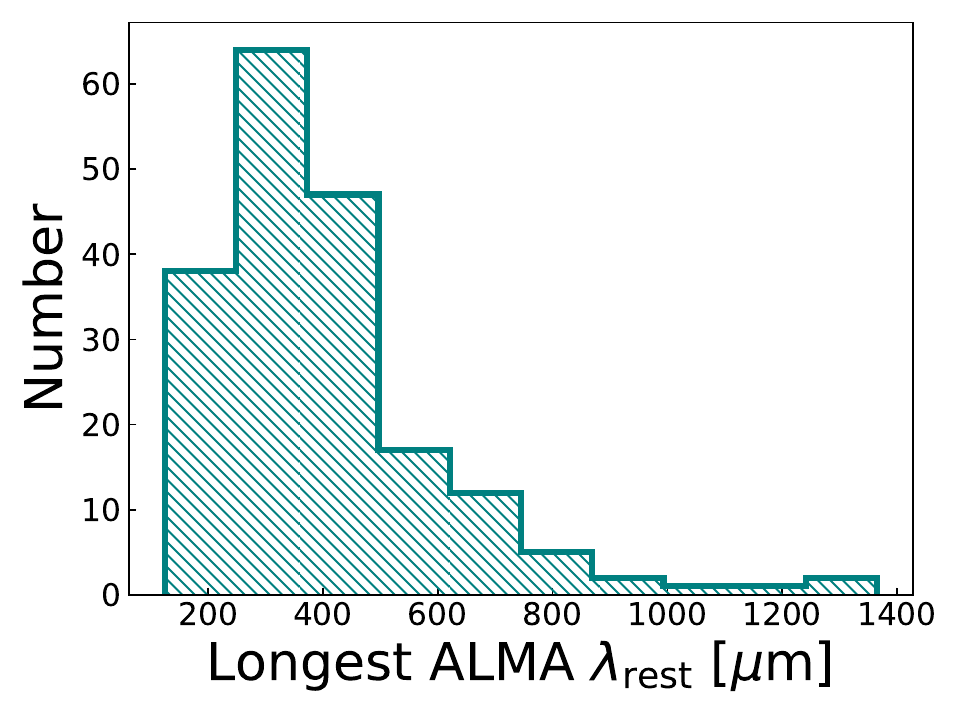}}
\caption{\small{Rest-frame wavelength corresponding to the longest observed ALMA band fluxes for each galaxy in the sample.}}
\label{fig:longest_restframe}
\end{figure}

 Once $\kappa_{\nu_0}$ and $\beta$ have been fixed, the main parameters that affect \md are the flux at the observed frequency $\nu_{\rm obs}$ and the rest-frame, mass-weighted, temperature $T_{\rm D}$. Several works suggest (or use) the dust temperature in a range between $\sim 15$K and $\sim 40$K \citep[see e.g.,][]{dunne2011dmd,dale2012dust,magnelli2014dust,pozzi2020dmd,pozzi2021dmd}. As it can be inferred from Eq. \ref{eq:mdust}, lower dust temperatures correspond to higher dust masses and vice versa, leading to uncertainties between $25\%$ and $50\%$ for \md with a different $T_{\rm D}$ assumption \citep{magnelli2020dmd}. Here we choose to use two values of rest-frame temperature, namely $T_{\rm D} = 25$K and $T_{\rm D} = 35$K, that encompass the range of possible values for the mass-weighted $T_{\rm D}$, supported by observations and models \citep[see e.g.,][]{magnelli2014dust,sommovigo2020tdust}{}{}. \cite{magnelli2014dust} found indeed that only a small fraction of galaxies with high sSFR have a $T_{\rm D} > 35$K.
To estimate \md from the observed flux, we choose to use the ALMA flux at the longest available wavelength ($S_{\rm ALMA, long}$). In our sample, the longest observed ALMA band always correspond to rest-frame values above $\lambda \sim 200 \mu$m, enabling us to to probe the R-J tail for every galaxy in our sample (see Figure \ref{fig:longest_restframe}).

\begin{figure}[]
\centering
{\includegraphics[width=.5\textwidth]{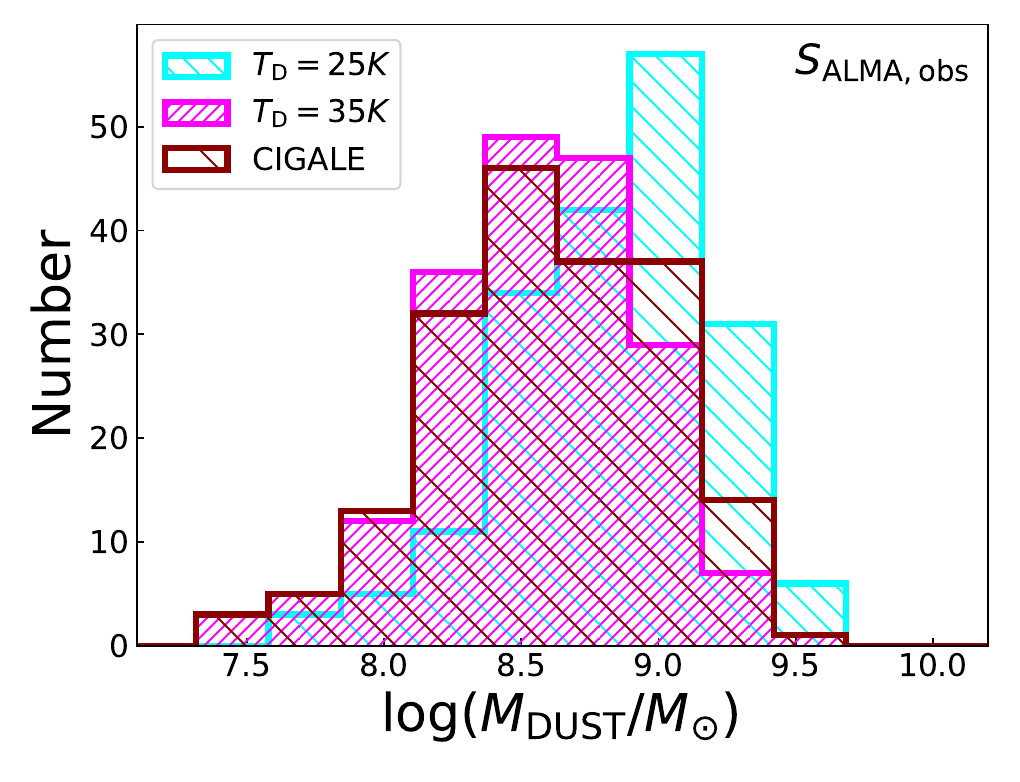}}
\caption{\small{Dust mass distribution as inferred using our three different approaches, i.e., the longest observed ALMA fluxes and assuming a single MBB with a dust temperature of 25K (cyan histogram) and 35K (magenta histogram). The darkred distribution corresponds to the dust masses given as an output of the \texttt{CIGALE} SED fitting.}}
\label{fig:mdust_comp_cigale}
\end{figure}

In Figure \ref{fig:mdust_comp_cigale}, we show the distributions for \md computed with the two different methods (SED-fit or single MBB) and with two different temperatures of the MBB (25K and 35K). 
The median values of the 25K and 35K dust masses are $7.6_{-4.7}^{+8.6} \times 10^{8}$ M$_{\odot}$ (hereafter $M_{\rm D,25K}$) and $3.7_{-2.1}^{+4.8} \times 10^{8}$ M$_{\odot}$ (hereafter $M_{\rm D,35K}$).
In this case the 25K \md is a factor $\sim 2$ larger than the 35K one. Moreover, as expected from the $T_{\rm D}$ distribution of the \texttt{CIGALE} SEDs (see Fig. \ref{fig:tdust_cigale}), the $M_{\rm D, 35K}$ and $M_{\rm D, CIGALE}$ are in very good agreement, with a difference of only $\sim 0.04$ dex in their median values.


\par In the rest of our analysis, we decided to derive the DMF and DMD using these three different dust mass estimates, i.e., $M_{\rm D, cigale}$, $M_{\rm D, 25K}$ and $M_{\rm D, 35K}$. $M_{\rm D, cigale}$ can be considered as our fiducial estimates because it is independent of the assumption made on the dust temperature and of the reference wavelength of the monochromatic flux. $M_{\rm D, 35K}$ should provide very consistent measurement but has the advantage compare to $M_{\rm D, cigale}$ to solely rely on the observed mm flux without the need for the UV-to-IR energy balance assumption made in \texttt{CIGALE}. Finally, while $M_{\rm D, 25K}$ seem to be disfavored in Fig. \ref{fig:mdust_comp_cigale}, we decided to keep it as the 25K assumption is commonly made in the literature \citep{magnelli2020dmd,pozzi2021dmd}.

%
%
\section{Dust mass function and dust mass density}\label{sec:DMFD} 

\subsection{$V_{\rm MAX}$ method and the dust mass function} \label{subsec:vmax_method}
In order to derive the DMF, we applied the \cite{Schmidt1970vmax} “maximum comoving volume” ($V_{\rm MAX}$) method that, based on the data, allowed us to derive the observed DMF without making any assumption relative to its shape. The areal coverage as a function of the flux has already been derived by \cite{Traina2024sfrd} for the \a3 survey at a reference observed wavelength of 1200$\mu$m. This relation between cumulative area and limiting fluxes is here used to associate an accessible area above a certain flux with each source.
\par We derive the DMF in eight redshift bins from $z\sim 0.5$ to $z\sim 6$ and into dust mass bins of 0.5 dex width from log($M_{\rm D}/\rm M_{\odot})=6$ to log($M_{\rm D}/\rm M_{\odot})=10$. For each source in a $z$-$M_{\rm D}$ bin, its contribution to the DMF is obtained as follows. At first, we shift in redshift and $K$-correct its best-fit SED from the lower to the upper boundary of the corresponding redshift bin and compute its flux at 1200 $\mu$m. Then, we use this flux to infer the corresponding areal coverage at each $dz$ (i.e., the redshift step used to shift the SED of each source) by interpolating the areal coverage vs flux density curve derived in \cite{Traina2024sfrd}. Lastly, we combine the effective area obtained in this way with the element of volume at each redshift step and obtaine a co-moving volume over which a given source is accessible:
\begin{equation}
    V_{\rm MAX} = V_{\rm zmax} - V_{\rm zmin}, 
\end{equation}
where $V_{\rm zmax}$ and $V_{\rm zmin}$ are the sum of the subvolume in each $dz$ shell up to the upper and lower limits of the bin, respectively. In particular, $V_{\rm zmax}$ can either be the volume at the upper bound of each $dz$ bin or the maximum volume reachable by considering the $S/N$ limit of the survey (i.e., corresponding to the $z$ at which the area would be zero).
Finally, we corrected the $V_{\rm MAX}$ by taking into account the completeness and spuriousness corrections derived by \cite{liu2019a31}, and we obtained the $\Phi(M,z)$ by summing each $1/V_{\rm MAX}$ in a certain luminosity-redshift bin. The completeness threshold is computed by rescaling all the observed 1200$\mu$m fluxes of each SED to the faintest observed 1200 $\mu$m flux density in this redshift bin and then taking the highest $M_{\rm D}$ of these SEDs. This latter value represents the $M_{\rm D}$ below which our sample is not 100$\%$ complete. We derived the DMFs using either the dust mass inferred from \texttt{CIGALE} or from the ALMA observed flux at the longest wavelength available and assuming $T_{\rm D} = 25$K, $T_{\rm D}=35$K. In each case we divided the sample in eight, similarly populated, redshift bins (0.5-1.0; 1.0-1.5; 1.5-2.0; 2.0-2.5; 2.5-3.0; 3.0-3.5; 3.5-4.5; 4.5-6.0). Fig. \ref{fig:25_35_a3only} shows the three dust mass functions derived with the different assumptions (i.e., DMF$_{\rm cigale}$, DMF$_{\rm 25K}$, DMF$_{\rm 35K}$) and the data points of our fiducial DMF are reported in Table \ref{tab:dmf}.
\par We compare our DMF data points with the few estimates available in similar redshift ranges from the literature. In particular, at $0.5 < z < 2.5$, we compare the \a3 DMFs with the results by \citet{pozzi2020dmd}. Firstly, it is important to underline the main differences between the two samples. The {\it Herschel} data sample is much larger than the \a3 sample ($\sim 6000$ and $\sim 190$ galaxies, respectively), leading to lower uncertainties on the estimated DMFs. Moreover, the {\it Herschel} sample traces a wider range of dust masses, both at the faint (due to a better sensitivity) and the bright-ends (larger co-moving volume probed). In the redshift and luminosity range in common, in the \a3 and the {\it Herschel} results, we find a weak agreement with the DMF$_{\rm 25K}$ and a much better consistency when using the $35$K dust masses, as well as the \texttt{CIGALE} $M_{\rm D}$. This agreement could be explained by the method used in \citet{pozzi2020dmd} to derive $T_{\rm D}$, which considers the dust temperature as a function of the redshift and the specific star formation rate \citep{magnelli2014dust}, which leads to $T_{\rm D} > 25$K (thus lower \md values), increasing with redshift up to $\sim 35-40$ K at $z \sim 2$. At higher redshifts, very few works have investigated the DMF. In Figure \ref{fig:25_35_a3only} and \ref{fig:cigale_herschel}, we report the \citet{dunne2003dmf} DMF at $1<z<5$ (obtained from a SCUBA sample of dusty galaxies), the point by \citet{magnelli2019dmd} at $3.1<z<4.6$, derived using the IRAM/GISMO 2mm Survey in the COSMOS field, and the ALPINE DMF at $z \sim 4.5$ \citep{pozzi2021dmd}. Within the large uncertainties of our data points, the three DMF estimates agree with the best-fit by \cite{dunne2003dmf}. While the DMF$_{\rm 35K}$ and DMF$_{\rm CIGALE}$ agree very well with the data point by \cite{magnelli2019dmd} at $z \sim 3.5$, the DMF$_{\rm 25K}$ is weakly consistent within the errors. Finally, our DMF$_{\it 25}$, as expected, is in a good agreement with the ALPINE DMF estimate, derived assuming $T_{\it D} = 25$K.


\begin{table*}[]
\centering

\renewcommand{\arraystretch}{1.5}
\caption{Dust mass function inferred from the \a3 database.}
\begin{threeparttable}
\begin{tabular}{ccccccccc}
\hline
\hline
log($M_{\rm D, CIGALE}/M_{\odot})$ & \multicolumn{4}{c}{log($\Phi/\rm Mpc^{-3} \rm dex^{-1}$)}                                                                                     \\ \hline
                        & $0.5<z  \le 1.0$                           & $1.0<z \le 1.5$                            & $1.5<z \le 2.0$                            & $2.0< z \le 2.5$                            \\ \cline{2-5} 
7.75-8.25             &                                            & (\textit{-4.68}$_{-0.35}^{+0.22}$)        &                                            &                                                         \\
\textbf{8.00-8.50}    & (\textbf{-4.46}$_{-0.86}^{+0.54}$)       & \textbf{(-4.68}$_{-0.36}^{+0.22}$)        & \textbf{(-3.73}$_{-0.42}^{+0.29}$)       & \textbf{(-3.90}$_{-0.40}^{+0.30}$)                    \\
8.25-8.75             & (\textit{-3.98}$_{-0.71}^{+0.45}$)         & (\textit{-3.68}$_{-0.37}^{+0.24}$)        & (\textit{-3.40}$_{-0.31}^{+0.18}$)       & \textit{-3.56}$_{-0.32}^{+0.19}$                    \\
\textbf{8.50-9.00}    & \textbf{-4.15}$_{-0.89}^{+0.55}$        & \textbf{-3.65}$_{-0.41}^{+0.25}$          & \textbf{-3.59}$_{-0.31}^{+0.19}$         & \textbf{-3.62}$_{-0.30}^{+0.17}$                      \\
8.75-9.25             &                                            & \textit{-4.46}$_{-0.75}^{+0.41}$          & \textit{-3.92}$_{-0.35}^{+0.22}$         & \textit{-3.95}$_{-0.36}^{+0.21}$                     \\
\textbf{9.00-9.50}    &                                            & \textbf{-4.65}$_{-0.88}^{+0.56}$          & \textbf{-4.04}$_{-0.40}^{+0.24}$         & \textbf{-4.51}$_{-0.61}^{+0.39}$                     \\
9.25-9.75             &                                            &                                             & \textit{-4.68}$_{-0.87}^{+0.60}$         & \textit{-5.21}$_{-1.21}^{+0.82}$                     \\ \hline
\multicolumn{1}{l}{}    & $2.5<z \le 3.0$                            & $3.0<z \le 3.5$                            & $3.5<z \le 4.5$                            & $4.5\textless{}z \le 6.0$ \\ \cline{2-5} 
7.75-8.25             &                                            &                                            &                                            & (\textit{-4.84}$_{-1.19}^{+0.85}$)              \\
\textbf{8.00-8.50}    & \textbf{(-4.05}$_{-0.34}^{+0.23}$)   & \textbf{(-4.19}$_{-0.71}^{+0.42}$)   & \textbf{(-4.33$_{-0.44}^{+0.28}$)}   & \textbf{(-4.59}$_{-0.70}^{+0.44}$)               \\
8.25-8.75             & \textit{-3.91}$_{-0.41}^{+0.28}$  & \textit{-3.77}$_{-0.42}^{+0.29}$   & \textit{-4.11}$_{-0.41}^{+0.29}$   & (\textit{-4.78}$_{-0.65}^{+0.45}$)               \\
\textbf{8.50-9.00}    & \textbf{-3.84}$_{-0.28}^{+0.20}$    & \textbf{-3.73}$_{-0.38}^{+0.27}$     & \textbf{-4.27}$_{-0.39}^{+0.28}$     & (\textbf{-4.83}$_{-0.61}^{+0.46}$)                  \\
8.75-9.25             & \textit{-3.90}$_{-0.32}^{+0.22}$    & \textit{-3.95}$_{-0.40}^{+0.27}$    & \textit{-4.56}$_{-0.51}^{+0.37}$    & \textit{-4.77}$_{-0.71}^{+0.43}$                  \\
\textbf{9.00-9.50}    & \textbf{-4.64}$_{-0.65}^{+0.44}$    & \textbf{-4.39}$_{-0.62}^{+0.46}$    & \textbf{-4.84}$_{-0.67}^{+0.49}$     & \textbf{-5.14}$_{-0.92}^{+0.59}$                  \\
9.25-9.75             & \textit{-5.02}$_{-0.87}^{+0.62}$     & \textit{-5.08}$_{-1.17}^{+0.88}$     & \textit{5.01}$_{-0.91}^{+0.62}$      & \textit{-5.42}$_{-1.22}^{+0.90}$                  \\
\textbf{9.50-10.00}   &                                            &                                            &                                            & \textbf{-5.42}$_{-1.22}^{+0.90}$                  \\ 
9.75-10.25            &                                            &                                            &                                            &                                                       \\
\textbf{10.00-10.50}  &                                            &                                            &                                            & \textit{-5.39}$_{-1.20}^{+0.88}$                                                       \\
10.25-10.75           &                                            &                                            &                                            & \textbf{-5.39}$_{-1.20}^{+0.88}$                                                     \\ \hline \\
\end{tabular}
\begin{tablenotes}
   \small{ \item[*]{Bold (or italic) values represent independent mass bins. Values in brackets indicate mass bins that are below the completeness limit.}}  
\end{tablenotes}
\end{threeparttable}
\label{tab:dmf}

\end{table*}

\subsection{The MCMC fitting analysis} \label{subsec:dust_mass_function}
To obtain the best-fit parameters characterizing the DMF at different redshifts, we performed a Markov chain Monte Carlo (MCMC) fitting analysis, modeling the DMF with a Schechter function \citep[that has been found to be a better parametrization than the modified-Schechter, often used to fit the IR-LF, ][]{pozzi2020dmd}:
\begin{equation}
    \Phi(M_{\rm D}){\rm dlog}M_{\rm D} = \Phi^*\left(\frac{M_{\rm D}}{M^*_{\rm D}}\right)^{\alpha} {\rm exp}\left[-\frac{M_{\rm D}}{M^*_{\rm D}}\right]{\rm dlog}M_{\rm D},
\end{equation}
where $\alpha$ is the slope of the faint end and $M^*_{\rm D}$ and $\Phi^*$ represent the dust mass and normalization at the knee, respectively. \cite{pozzi2020dmd} found also an evolutionary trend with redshift for these parameters, expressed as:

\begin{equation}
\begin{cases}
  \Phi^* = \Phi^*_{0} (1+z)^{k_{\rm \rho1}} ~~~~~~~~~~~~~~~~~~~~~~~~ z < z_{\rm \rho0}\\
  \Phi^* = \Phi^*_{0} (1+z)^{k_{\rm \rho2}} (1+z_{\rm \rho0})^{(k_{\rm \rho1}-k_{\rm \rho2})}~z > z_{\rm \rho0},
\end{cases}
\end{equation}

\begin{equation}
\begin{cases}
 M^*_{\rm D} = M^*_{D,0} (1+z)^{k_{\rm l1}} ~~~~~~~~~~~~~~~~~~~~~~~ z < z_{\rm l0}\\
 M^*_{\rm D} = M^*_{D,0} (1+z)^{k_{\rm l2}} (1+z_{\rm l0})^{(k_{\rm l1}-k_{\rm l2})}~~ z > z_{\rm l0},
\end{cases}
\end{equation}
where $\Phi^*_{0}$ and $M^*_{D,0}$ are the normalization and characteristic dust mass at $z=0$ and $k_{\rm \rho1}$, $k_{\rm \rho2}$, $k_{\rm l1}$, and $k_{\rm l2}$ are the exponents for values lower and greater than $z_{\rm \rho0}$ and $z_{\rm l0}$ for $\Phi$ and $M_{\rm D}$, respectively. We thus performed an MCMC fit using simultaneously each point of the DMF associated with a redshift corresponding to the median redshift value of the underlying galaxy population in the bin.
\par We carried out the MCMC analysis using the \texttt{PYTHON} package \texttt{emcee} \citep{foremanmackey2013emcee}, using a set of 50 walkers and 10000 steps to explore the parameter space, discarding the first 1000 sampled draws of each walker (burnin). The log-likelihood was built in the following form:
\begin{equation}
    {\rm logL} = -\frac{1}{2} \sum \left(\frac{\Phi_{\rm Model} - \Phi}{\delta \Phi}\right)^2 .
\end{equation}

\begin{figure*} 
\centering
\includegraphics[width=0.7\textwidth]{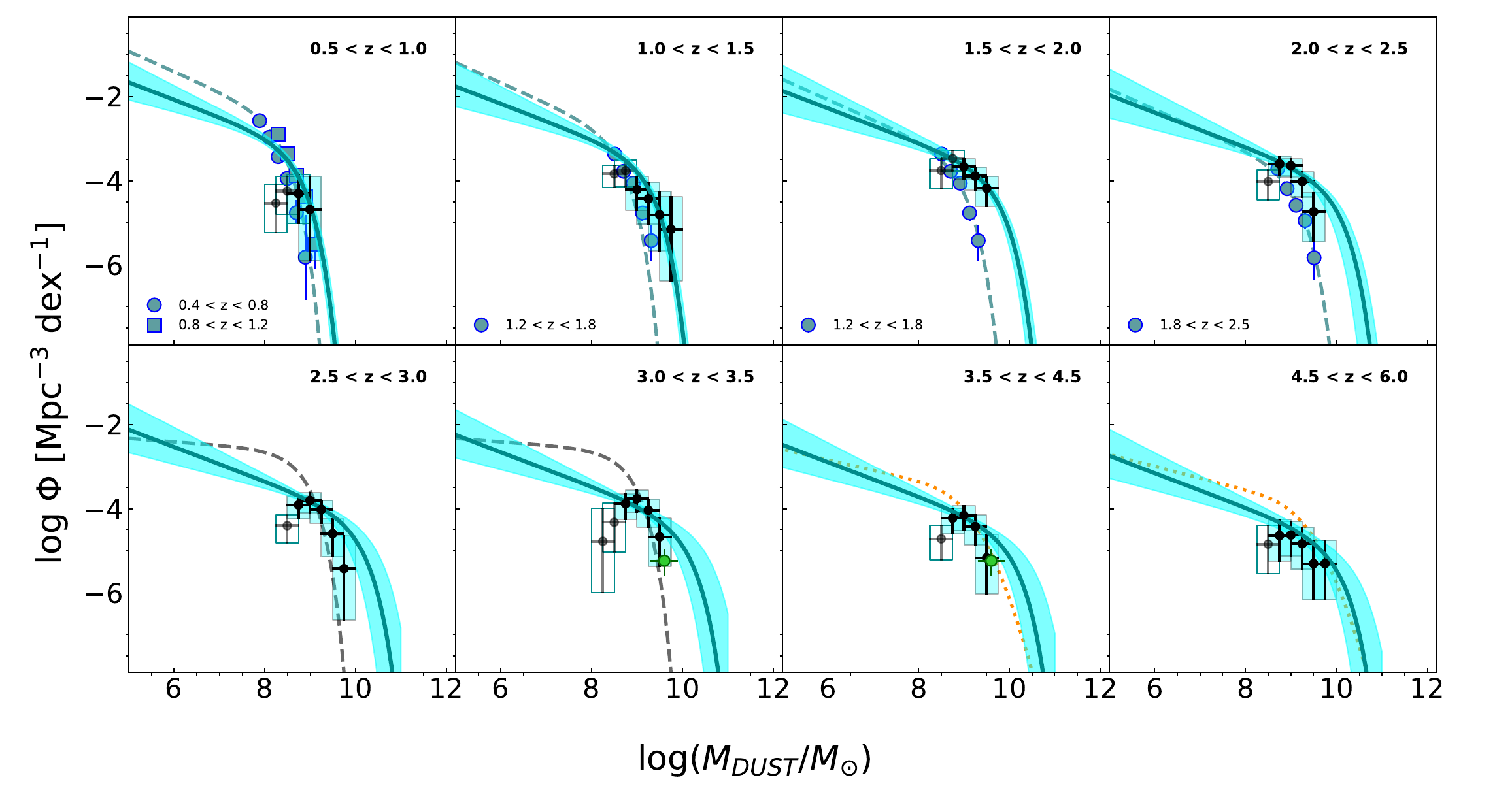}\\
\vspace{0.1cm} 
\centering
\includegraphics[width=0.7\textwidth]{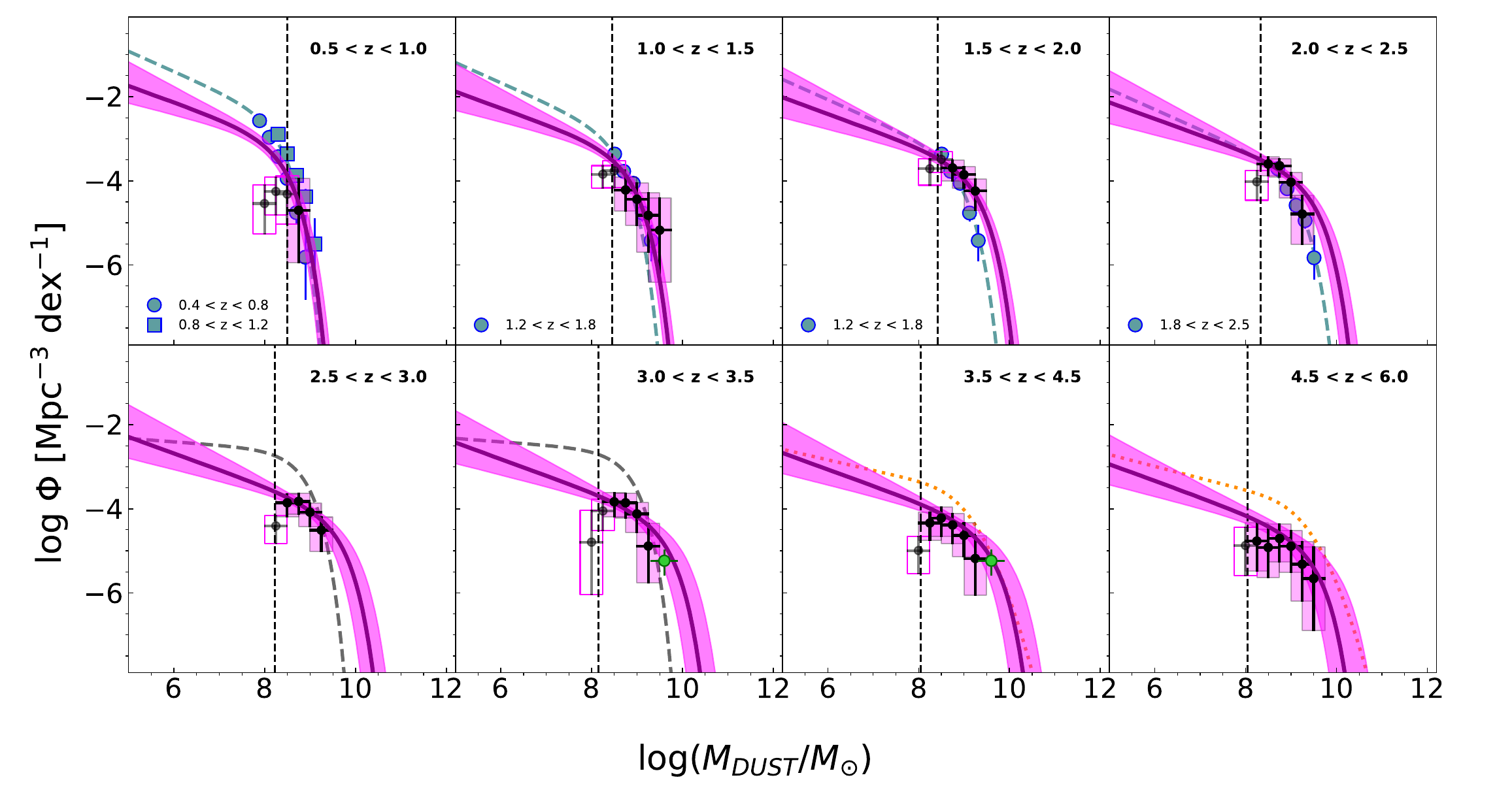}\\
\vspace{0.1cm} 
\centering
\includegraphics[width=0.7\textwidth]{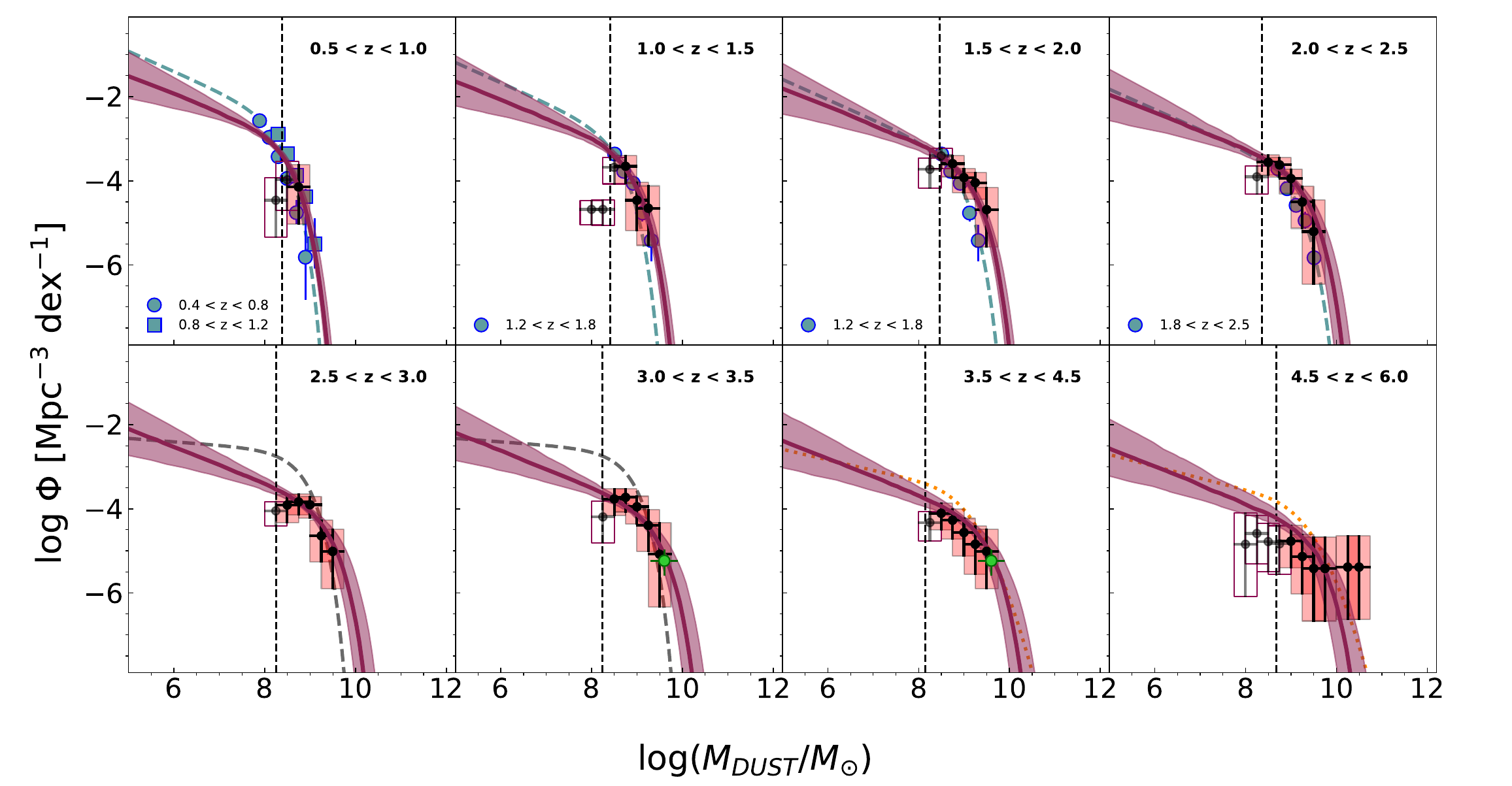}\\
\vspace{0.1cm} 
\centering
\includegraphics[width=0.7\textwidth]{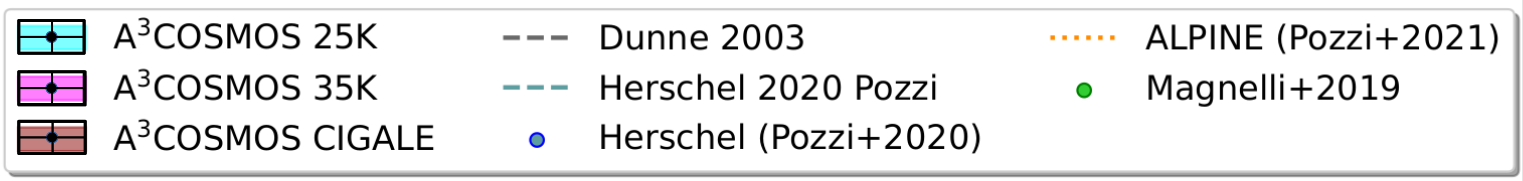}
\caption{\small{Dust mass function derived using the $V_{\rm MAX}$ method for dust masses of the galaxies computed assuming $T_{\rm D} =25$K (cyan squares and black circles with errors, computed following \citet{gehrels1986errors}, upper panel), $T_{\rm D} =35$K (magenta squares and black circles with errors, central panel) and using the dust mass from \texttt{CIGALE} (red squares and black circles with errors, lower panel). The best-fit, computed using the \a3 data points, is displayed as a red solid line, with shaded errorbands of the same colors. For this fit, $M_{\rm D}^*$, $\Phi_{\rm D}^*$ and $\alpha$ are free to vary.
For comparison, different estimates from the literature are reported. The light blue circles and squares are the values obtained by \cite{pozzi2020dmd} and the cyan dashed lines correspond to the best-fit. The grey dashed curves are from \cite{dunne2003dmf}. The blue circle is the estimate by \cite{magnelli2019dmd} and the dotted line is the ALPINE dust mass function by \cite{pozzi2021dmd}, assuming $T_{\rm D} = 25$K.}}
\label{fig:25_35_a3only}
\end{figure*}

\begin{figure*}
\includegraphics[width=1\textwidth]{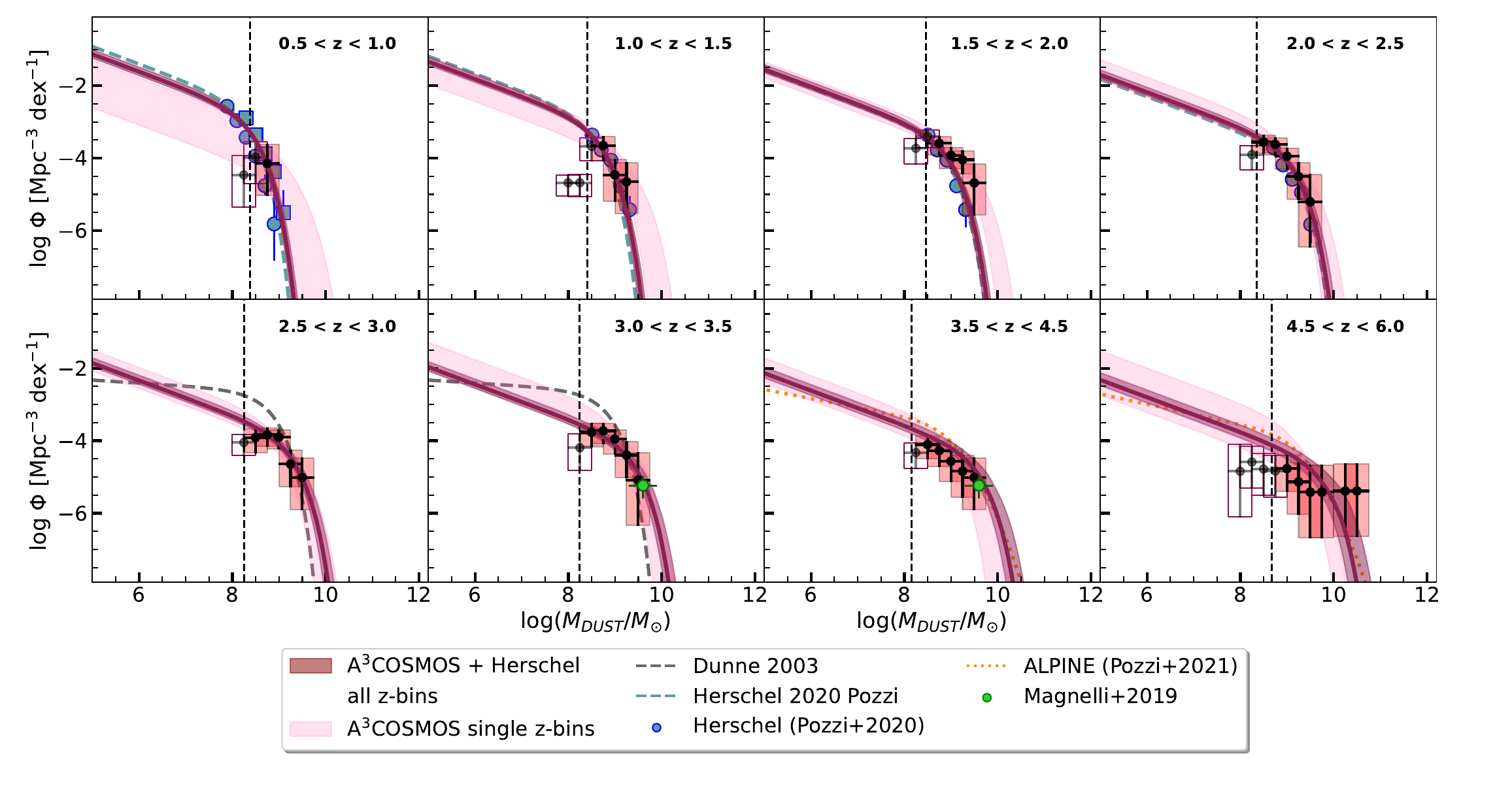}\\  
\caption{\small{Dust mass functions derived using the $V_{\rm MAX}$ method for dust masses of the galaxies obtained via SED fitting (red boxes and black circles with errors, computed following \citet{gehrels1986errors}). The evolutive best-fit of the \texttt{CIGALE} + {\it Herschel} is displayed as a red solid line, with shaded errorbands of the same colors, while the individual \a3 fit is shown as pink shaded area. Errors on the evolutive best-fit are computed by bootstrapping on the dust masses of the sources. For comparison, different estimates from the literature are reported. The light blue circles and squares are the values obtained by \cite{pozzi2020dmd} and the cyan dashed lines correspond to the best-fit. The grey dashed curves are from \cite{dunne2003dmf}. The blue circle is the estimate by \cite{magnelli2019dmd} and the dashed line is the ALPINE dust mass function by \cite{pozzi2021dmd}.}}
\label{fig:cigale_herschel}
\end{figure*}

With this procedure, we obtained a first best-fit of the three DMFs, by running the MCMC leaving the Schechter parameters (i.e, $\alpha$, $\Phi^*$ and $M^*$) free to vary (see Fig. \ref{fig:25_35_a3only}). For DMF$_{\rm 25K}$, $M^*_{\rm D}$ is significantly higher than what is found by \citet{pozzi2020dmd} and the best-fit has a very different shape. The DMF$_{\rm 35K}$ is instead more consistent with the \citet{pozzi2020dmd} estimate and the best-fit is also more in agreement, even though the $M^*_{\rm D}$ is still higher. Indeed, by fitting the DMF using only the \a3 data, we cannot robustly constrain all parameters of the Schechter function.
\par
\begin{figure}[]
\centering
{\includegraphics[width=.5\textwidth]{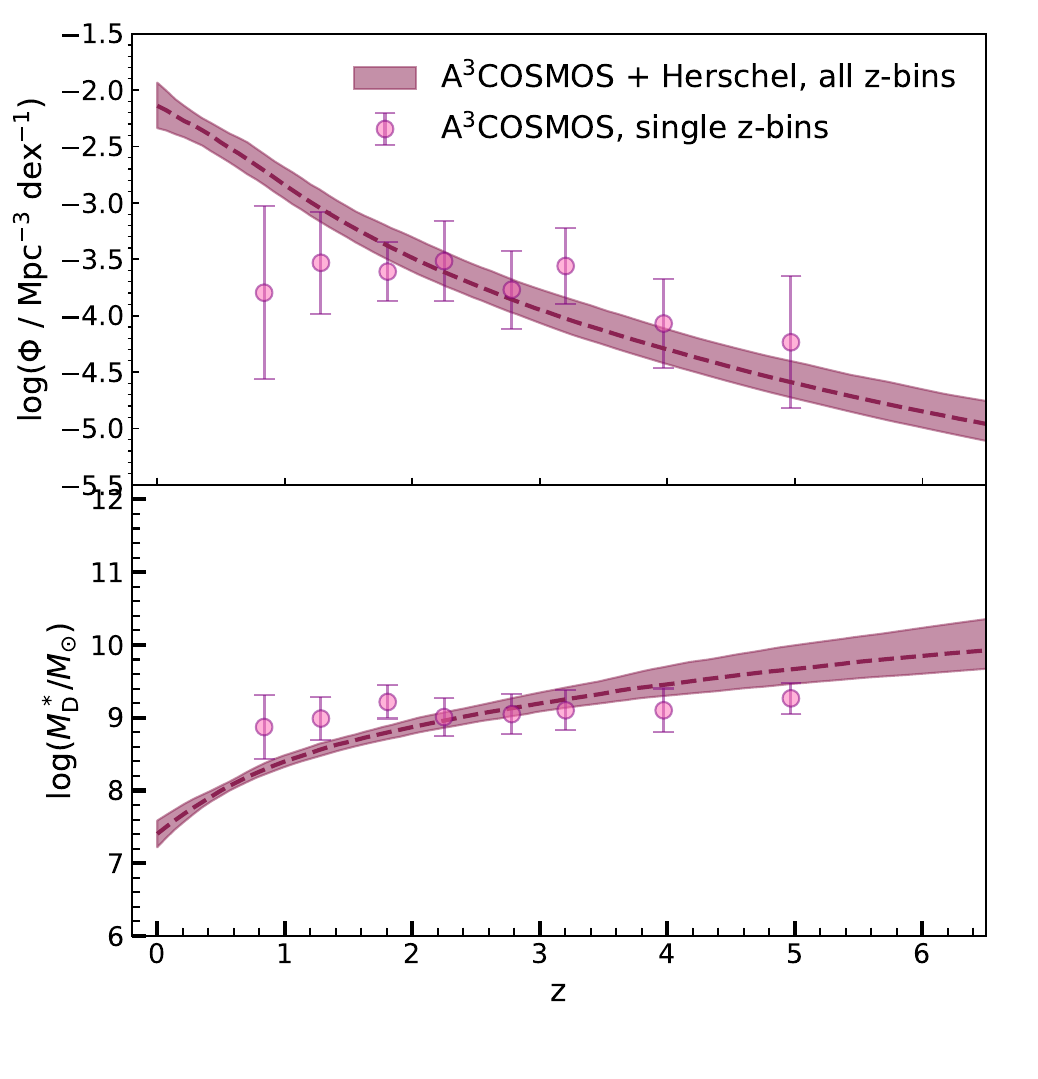}}
\caption{\small{Evolution with redshift of $M_{\rm D}^*$ and $\Phi^*$ obtained by fitting the \a3 and {\it Herschel} DMFs (red dashed lines and shaded areas), compared to the same parameters derived by fitting the \a3 \texttt{CIGALE} DMF in each redshift bin separately (pink circles).}}
\label{fig:evolution_stars}
\end{figure}

 To improve the quality of our fit, as already done by \citet{Traina2024sfrd} for deriving the LF best fit, we exploit the {\it Herschel} data derived by \citet{pozzi2020dmd} to fit a combined \a3 + {\it Herschel} DMF. The greater statistic of the {\it Herschel} data at $0 < z <2.5$, allows us to combine the accuracy of the DMF by \citet{pozzi2020dmd} at lower-mid redshifts, with the capability of ALMA to explore the dust content of galaxies in the high-$z$ Universe. With this combined datasets, we are able to better constrain the shape of the DMF and derive more reliable best fit parameters. The {\it Herschel} DMF was derived by assuming that the dust temperature is a function of $z$ and of the sSFR, as reported in \cite{magnelli2014dust}. This means that the galaxies in that sample do not have a fixed temperature, so it is not possible to coherently combine those estimates of the DMF with ours DMF$_{\rm 25K}$ or DMF$_{\rm 35K}$. For this reason, we decided to combine the \citet{pozzi2020dmd} DMF with the \texttt{CIGALE} DMF, since it does not assume a fixed temperature and is also more consistent with the \citet{pozzi2020dmd} data.
We ran the MCMC with $\alpha$ fixed to the values found by \citet{pozzi2020dmd} (i.e., $\alpha$=1.48), using flat prior distributions for the two free parameters ($\Phi^*$ and $M^*$), with log($\Phi^*_{0}$) between $-2.5$ and $-2$ and log($M^*_{\rm D,0}$) between 6 and 7.5. The results of this DMF fit are shown in Figure \ref{fig:cigale_herschel} and summarized in Table \ref{tab:dmf_mcmc_evol} and we consider this best-fit as our fiducial DMF. As a comparison, we also overplot the best-fit obtained considering the \a3 DMF data-points in each redshift bin separately. This fit shows larger errors due to the lower number of data and the Schechter parameters also differ from what we obtained with the evolutive fit. The typical dust mass at each redshift bin (i.e., $M_{\rm D}^*$), from our fiducial best-fit, is almost constantly increasing by $\sim 1$ dex, from $z \sim 0.5$ to $z \sim 6$. This means that, as for $L_{\rm IR}^*$, massive galaxies are typically more dust rich at higher redshifts than their local counterparts. Similarly, the density $\Phi^*$ decreases steeply towards higher redshifts (by $\sim 2$ dex). Comparing this values with the individual fit ones (see Figure \ref{fig:evolution_stars}), we note that without the combination with the {\it Herschel} data, the typical $M_{\rm D}^*$ is nearly constant, while the density ($\Phi^*$) decreases at $z>3$.

\begin{table*}[]
\centering

\renewcommand{\arraystretch}{1.5}
\caption{Best-fit parameters at the knee of the DMF inferred by combining the \a3 dataset with the {\it Herschel} data points.}
\begin{threeparttable}
\begin{tabular}{ccccccc}
\hline
\hline
$z$       & \begin{tabular}[c]{@{}c@{}}log($M^*/M_{\odot})$ \\ 16th\end{tabular} & \begin{tabular}[c]{@{}c@{}}log($M^*/M_{\odot})$ \\ 50th\end{tabular} & \begin{tabular}[c]{@{}c@{}}log($M^*/M_{\odot})$\\  84th\end{tabular} & \begin{tabular}[c]{@{}c@{}}log($\Phi^*/\rm Mpc^{-3} \rm dex^{-1}$)\\  16th\end{tabular} & \begin{tabular}[c]{@{}c@{}}log($\Phi^*/\rm Mpc^{-3} \rm dex^{-1}$)\\  50th\end{tabular} & \begin{tabular}[c]{@{}c@{}}log($\Phi^*/\rm Mpc^{-3} \rm dex^{-1}$)\\  84th\end{tabular} \\ \hline
$0.5-1.0$ & 8.22                                                                & 8.29                                                                & 8.37                                                                & -2.84                                                                      & -2.71                                                                & -2.57                                                                      \\
$1.0-1.5$ & 8.48                                                                & 8.56                                                               & 8.65                                                                & -3.17                                                                      & -3.05                                                                & -2.89                                                                      \\
$1.5-2.0$ & 8.71                                                                & 8.80                                                                & 8.89                                                                & -3.50                                                                      & -3.38                                                               & -3.21                                                                      \\
$2.0-2.5$ & 8.86                                                                & 8.96                                                               & 9.07                                                                & -3.73                                                                      & -3.61                                                                 & -3.43                                                                      \\
$2.5-3.0$ & 9.02                                                                & 9.13                                                                & 9.27                                                                & -3.97                                                                      & -3.86                                                                & -3.68                                                                      \\
$3.0-3.5$ & 9.14                                                                & 9.25                                                                & 9.41                                                                & -4.15                                                                     & -4.03                                                                & -3.84                                                                       \\
$3.5-4.5$ & 9.30                                                                & 9.45                                                                & 9.70                                                                & -4.42                                                                      & -4.29                                                               & -4.11                                                                     \\
$4.5-6.0$ & 9.46                                                                & 9.66                                                               & 9.99                                                                & -4.72                                                                        & -4.59                                                               & -4.40  \\ \hline                                                                   
\end{tabular}
\begin{tablenotes}
   \small{ \item[*]{Dust masses (M$^*$) and normalizations ($\Phi^*$) with 16th, 50th, and 84th percentiles at the knee in the eight redshift bins obtained through the MCMC analysis, in the ALMA+{\it Herschel} case, using the information from all the redshifts together.}}  
\end{tablenotes}
\end{threeparttable}
\label{tab:dmf_mcmc_evol}
\end{table*}

\subsection{Dust mass density}
\begin{figure*}
\includegraphics[width=1\textwidth]{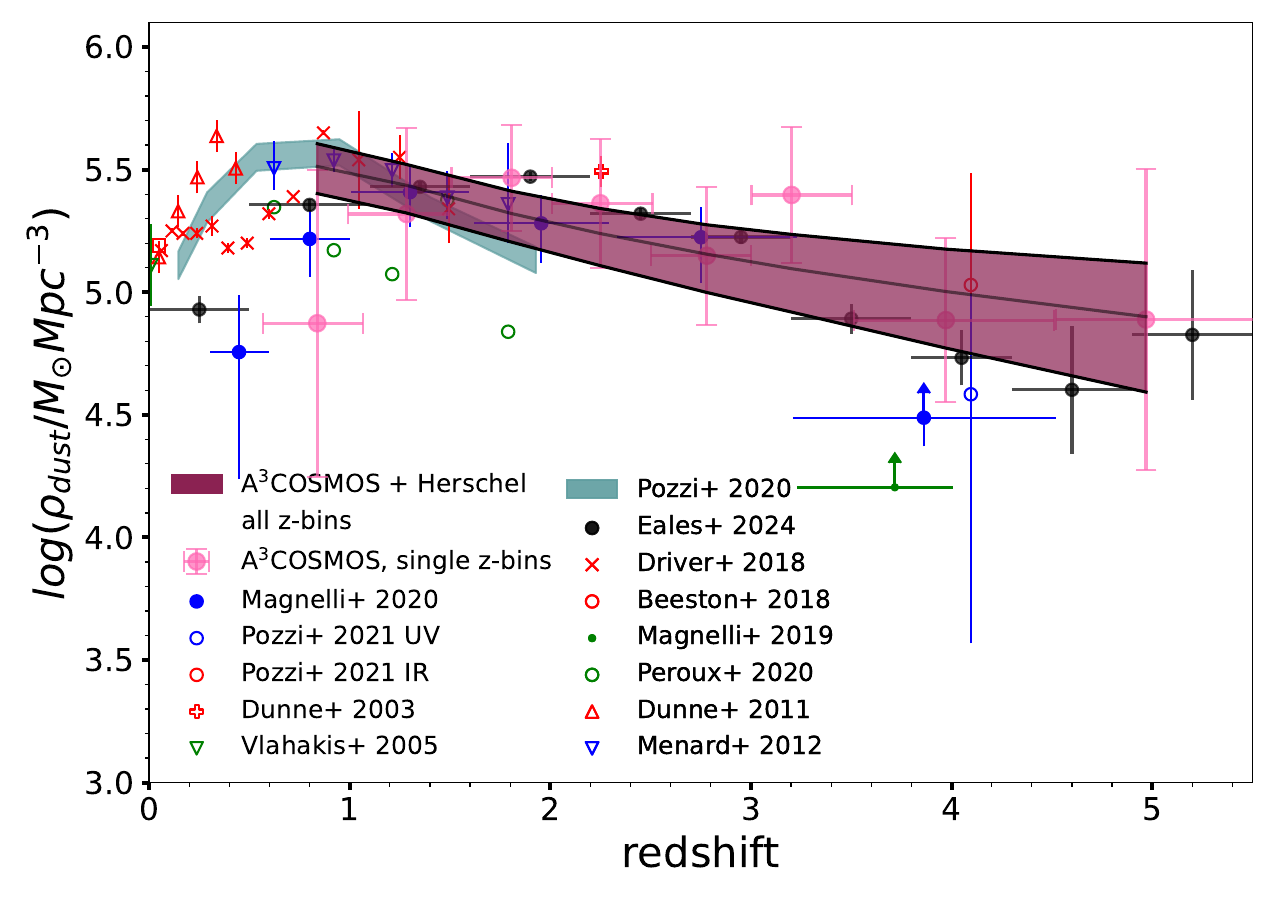}\\  
\caption{\small{Dust mass density evolution with redshift, derived by integrating the \texttt{CIGALE} + {\it Herschel} dust mass function in each redshift bin (red shaded area). The pink circles, with errors, represent the DMD obtained by fitting the DMF individually at each redshift bin.
The red plus marker shows the values by \cite{dunne2003dmf}; the green triangles are the estimates by \cite{vlahakis2005dmd}; the red triangles are the DMD points by \cite{dunne2011dmd}; the blue triangles indicates the estimate by \cite{menard2012dmd}; the red crosses are the data points by \cite{driver2018dmd}; the red circles are the data by \cite{beeston2018dmd}; the estimate by \cite{magnelli2019dmd} is shown as a green filled circle; the green empty circles are the points by \cite{peroux2020dmd}; the estimates by \cite{pozzi2021dmd} are shown as blue and red circles; the dust mass densities from \cite{magnelli2020dmd} are displayed as blue filled circles; the black points are the DMD estimates by \cite{eales2024dmd_herschel}; finally, the light blue shaded area represents the estimate of the dust mass density by \cite{pozzi2020dmd}. For a self-consistent comparison, we rescaled the data by \cite{magnelli2020dmd} and \cite{pozzi2021dmd} to a $T_{\rm D} = 35$K.}}
\label{fig:dmd_cigale}
\end{figure*}

By integrating the DMF best-fit (at $4<{\rm log}(M_{\rm D} / M_{\odot})<11$) in each redshift bin, we can measure the amount of dust, per unit of co-moving volume, obtaining the DMD. Figure \ref{fig:dmd_cigale} shows the DMD obtained by integrating the combined \a3 \texttt{CIGALE} and {\it Herschel} DMF best-fit. We report the values of the DMD in Table \ref{tab:dmd_values}. 
To allow the comparison of the results by \cite{magnelli2020dmd} and \cite{pozzi2021dmd} with ours, we rescaled their DMDs from $T = 25$K to $T=35$K (i.e., $M_{\rm D, 35K} = 0.54 \times M_{\rm D, 25K}$, see Section \ref{sec:mdust}). 
We find the DMD to be following the shape of the \cite{pozzi2020dmd} DMD, with a smoother decrease from $z =1$ to higher redshifts. Our results are in excellent agreement with the estimates by \cite{menard2012dmd}, obtained for MgII absorbers. We are also consistent with the DMD from \cite{driver2018dmd}, in which they studied galaxies from the GAMA, G10-COSMOS and 3D-HST surveys, between $z \sim 1$ and $z \sim 2$. We are consistent with \cite{magnelli2020dmd} at $z \sim 1.5-3$. The DMD by \cite{peroux2020dmd} is instead significantly lower between $z \sim 1.5$ and $z \sim 4$. The reason behind the discrepancy with our result may be ascribed to a possible selection bias in the estimate by \cite{peroux2020dmd} (also mentioned in their paper) toward dust-poor galaxies. Indeed their results are based on a sample of optically selected quasars, for which they derived the DMD combining the gas mass density and the dust-to-gas ratio. At $z \sim 4.5$, we are fully consistent with the results obtained from the rest-frame FIR selected galaxies in the ALPINE ALMA survey \citep{pozzi2021dmd}. Comparing our results with the recent estimates by \cite{eales2024dmd_herschel}, obtained using {\it Herschel}-ATLAS, we find a good agreement at $z \sim 1.5$, while our DMD is lower at $2 < z <3$ and slightly higher at $3<z<5$.

\begin{table}[]
\centering
\renewcommand{\arraystretch}{1.5}
\caption{Dust mass density obtained by integrating the DMF best-fit in our eight redshift bins, for the \texttt{CIGALE} \a3 + {\it Herschel} fit.}
\begin{threeparttable}

\begin{tabular}{cccc}
\hline
\hline
$z$       & \begin{tabular}[c]{@{}c@{}} \\ \\ 16th\end{tabular} & \begin{tabular}[c]{@{}c@{}}log$\rho_{\rm DUST}$ \\ {[}$\rm M_{\odot} \rm Mpc^{-3}${]} \\ 50th\end{tabular} & \begin{tabular}[c]{@{}c@{}} \\ \\ 84th\end{tabular} \\ \hline
$0.5-1.0$ & $5.40$                                                                                                 & $5.51$                                                                                                 & $5.61$                                                                                                 \\
$1.0-1.5$ & $5.32$                                                                                                 & $5.44$                                                                                                 & $5.52$                                                                                                 \\
$1.5-2.0$ & $5.20$                                                                                                 & $5.32$                                                                                                 & $5.41$                                                                                                 \\
$2.0-2.5$ & $5.11$                                                                                                 & $5.24$                                                                                                 & $5.34$                                                                                                 \\
$2.5-3.0$ & $5.00$                                                                                                 & $5.16$                                                                                                 & $5.27$                                                                                                 \\
$3.0-3.5$ & $4.92$                                                                                                 & $5.10$                                                                                                 & $5.24$                                                                                                 \\
$3.5-4.5$ & $4.77$                                                                                                 & $5.00$                                                                                                 & $5.18$                                                                                                 \\
$4.5-6.0$ & $4.59$                                                                                                 & $4.90$                                                                                                 & $5.12$    \\ \hline                                                                                            
\end{tabular}

\begin{tablenotes}
   \small{ \item[*]{The third column is the median value, while second and fourth columns show the lower and upper 16th boundaries.}}  
\end{tablenotes}
\end{threeparttable}
\label{tab:dmd_values}
\end{table}

%
%

\section{Discussion}\label{discussion}

\begin{figure}
\includegraphics[width=0.5\textwidth]{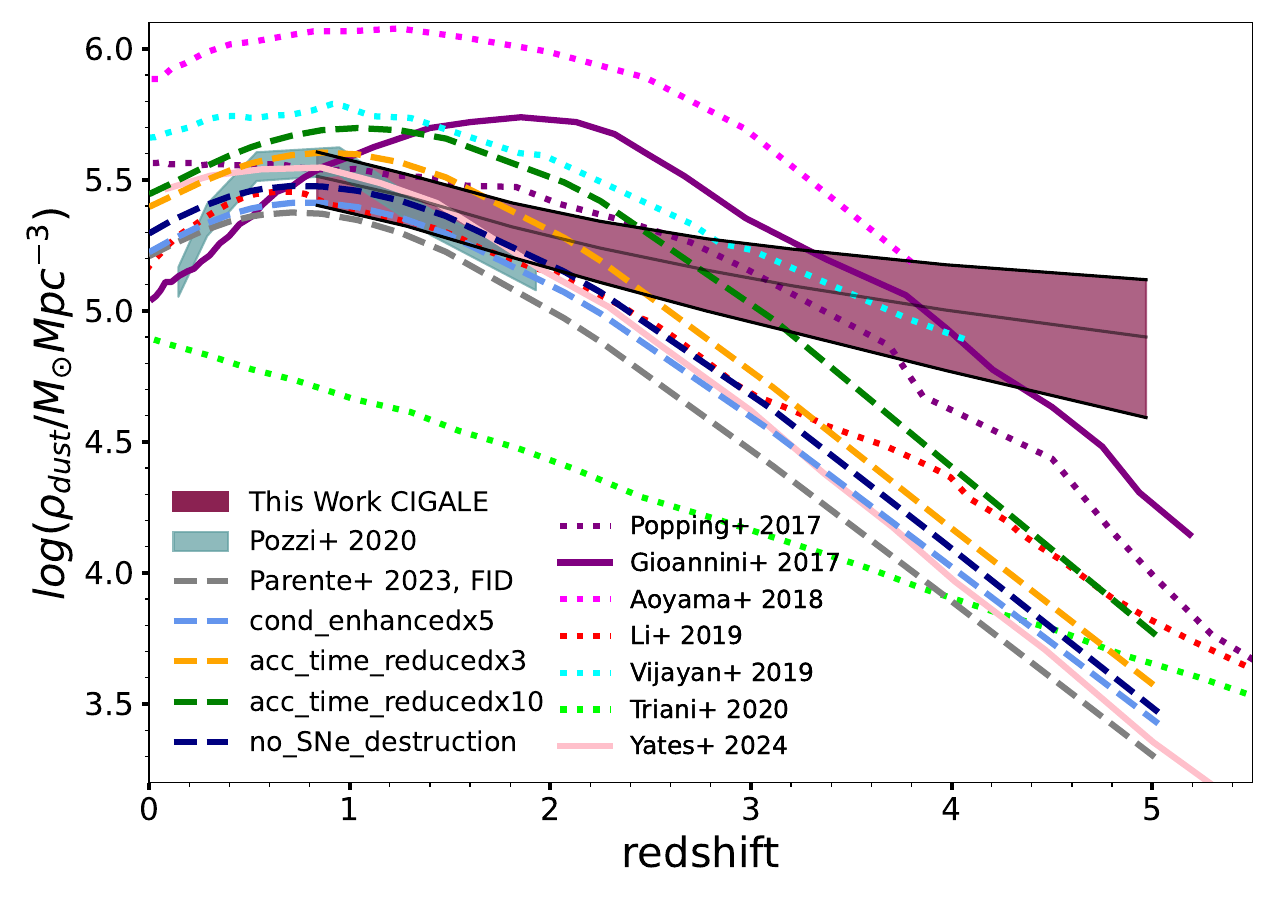}\\  
\caption{\small{Comparison between our DMD (derived using the \texttt{CIGALE} dust masses) and predictions from simulations. The purple solid line is the DMD by \cite{gioannini2017dmd_sim}, while the purple dashed line represents the prediction by \cite{popping2017dmd_sim}. The magenta dotted line is the prediction by \cite{aoyama2018dmd_sim}, the cyan dotted line is the DMD predicted by \cite{vijayan2019dmd_sim}, the light-green dotted curve is by \cite{triani2020dust_sim}, red dotted line is the prediction by \citet{li2019dmd_sim} and the pink solid line is the DMD from \citet{yates2024dust}. The grey, light blue, yellow, green and blue dashed lines are the DMDs estimated using the models by \cite{parente2023dmdm_sim} with different prescriptions.}}
\label{fig:dmd_models}
\end{figure}

\begin{figure}
\includegraphics[width=0.5\textwidth]{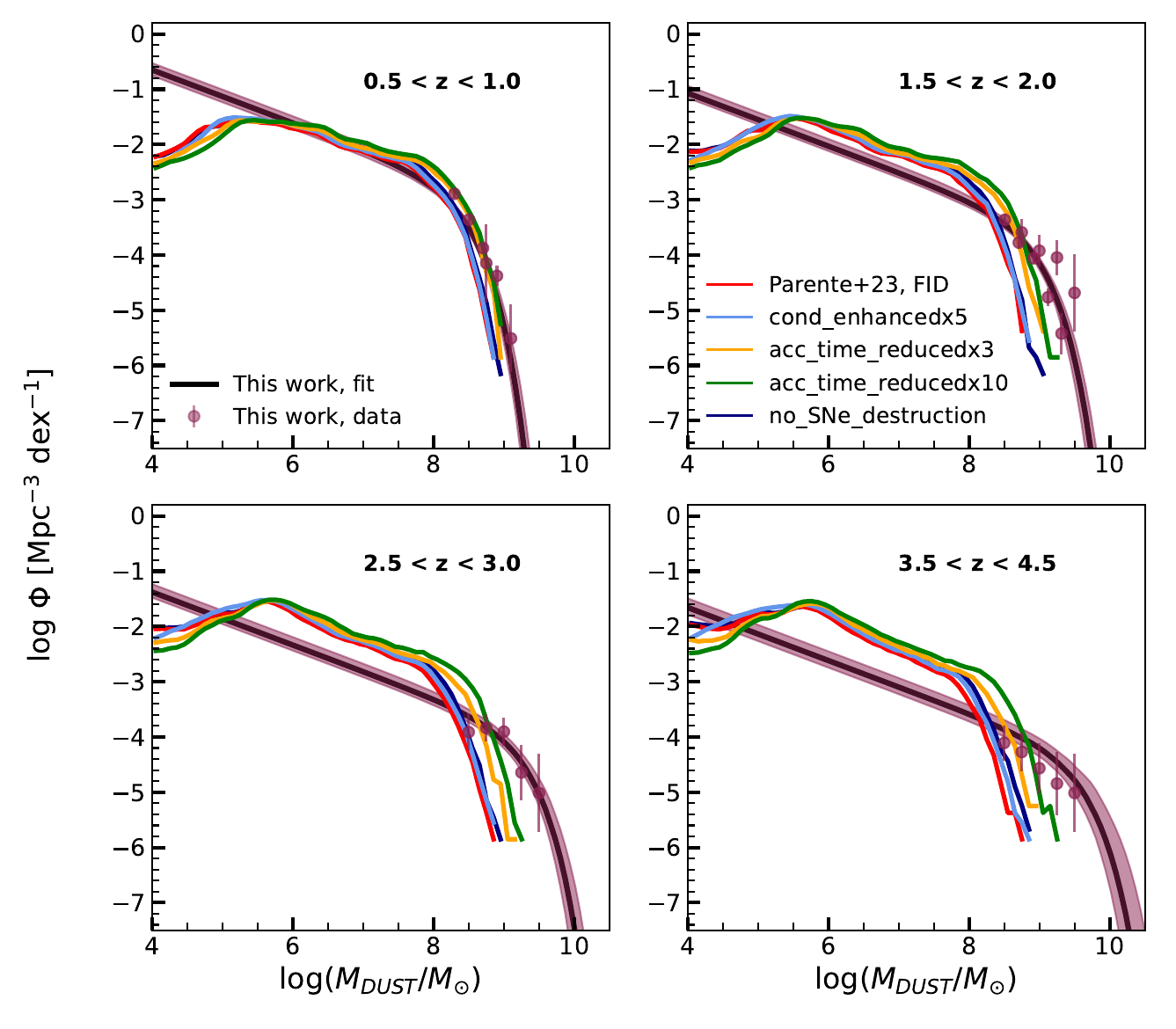}\\  
\caption{\small{Comparison between the DMF derived in this work (black solid line with shaded area and red points) and the predictions by \cite{parente2023dmdm_sim}, with different prescriptions.}}
\label{fig:dmf_parente}
\end{figure}

In this Section, we compare our results with predictions for the DMF and DMD from models in the literature. In particular, we consider the semi-analytical model by \cite{parente2023dmdm_sim} for comparison to the DMF, and the predictions by \cite{gioannini2017dmd_sim}, \cite{popping2017dmd_sim}, \cite{aoyama2018dmd_sim}, \citet{li2019dmd_sim}, \cite{vijayan2019dmd_sim}, \cite{triani2020dust_sim}, \cite{parente2023dmdm_sim} and \citet{yates2024dust} to compare with the DMD.

Although different in the details implementation, the dust models in the aforementioned works are conceptually similar. They all include: (i) dust production by stellar sources (SNe and AGB stars) (ii) dust evolution in the ISM, in particular accretion of gas-phase metals onto pre-existing grains, and destruction of grains in hostile environments (e.g. SN shocks and the hot phase) (iii) astration of grains, that is grains returning into newly formed stars. All these processes play a role in determining the amount of dust present in galaxies. Specifically, the evolution of dust within the ISM turns out to be very important. Indeed, simulations indicate that the bulk of the dust mass observed today originates from grain accretion within the ISM, while stellar production only contributes a small fraction ($\lesssim 10 \, \%$) to the overall dust budget (e.g. \citealt{vijayan2019dmd_sim}, \citealt{parente2023dmdm_sim}).\\

The DMD predicted by the aforementioned model is shown in Fig. \ref{fig:dmd_models}. There is a great dispersion, spanning about one order of magnitude at all redshifts. Also, there is no agreement in terms of shape. In some of the models (e.g. \citealt{triani2020dust_sim}, \citealt{popping2017dmd_sim}) we observe a $\rho_{\rm dust}$ increasing with cosmic time, while some others (e.g. \citealt{gioannini2017dmd_sim}, \citealt{aoyama2018dmd_sim}, \citealt{li2019dmd_sim}, \citealt{parente2023dmdm_sim}) predict a clear drop from $z\simeq 1-2$ to $z=0$, which is more in line with observations. In general, none of the models is able to match both the shape and normalization of the observed $\rho_{\rm dust}$.

The different performances among the various models in predicting cosmic dust abundance originate from distinct underlying reasons that are challenging to pinpoint. Firstly, while the theoretical frameworks of these models are conceptually similar, their practical numerical implementation can vary significantly. Secondly, a major contributing factor could be that the dust model is implemented on top of different galaxy evolution models, each incorporating sub-grid recipes for processes (e.g. star formation, chemical enrichment) which are crucial for the production and evolution of dust mass. Analyzing these disparities is not trivial and goes beyond the scope of the current work. Instead, we adopt an alternative approach, aiming to quantify the influence of specific dust-related mechanisms — namely stellar production, ISM accretion, and SN-driven destruction — on the overall dust budget. To do this, we standardize the galaxy evolution framework to accentuate the impacts of the targeted processes. Specifically, we conduct multiple simulations using the \cite{parente2023dmdm_sim} model\footnote{This is based on the \textsc{L-Galaxies} SAM \citep{henriques2020lgalaxies}. It is noteworthy that this model is able to reproduce the general trend of redshift evolution of the dust mass density within the uncertainties (see Fig 10 of \citealt{parente2023dmdm_sim}).} to explore potential ways for increasing the dust mass at the redshifts investigated here. The conducted simulations are as follows:

\begin{itemize}
    \item \textbf{cond\_enhancedx5}: the dust condensation efficiency is enhanced by a factor of $5$, to mimic a larger dust production by stellar sources (both AGBs and SNe);
    \item \textbf{acc\_time\_reducedx3}: the ISM grains accretion timescale is reduced by a factor of $3$, to mimic a more efficient accretion in molecular clouds;
    \item \textbf{acc\_time\_reducedx10}: same as before, but the timescale is reduced by a factor of $10$;
    \item \textbf{no\_SNe\_destruction}: grains destruction in SN shocks is switched off.
\end{itemize}

In Fig. \ref{fig:dmd_models}, we present the DMD results from these experiments. Switching off grains destruction by SNe only marginally increases the overall dust abundance compared to the fiducial run. This occurs because the metals generated by the SNe-driven destruction rapidly recombine into dust grains through the highly efficient accretion process. Similarly, increasing the AGBs and SNe production of grains has a minor effect on the DMD, raising it by a factor of $\simeq 1.4$ at $z \gtrsim 1$. Conversely, increasing the efficiency of grains accretion has a stronger impact on the DMD, boosting it by a factor of $\gtrsim 2$ compared to the fiducial model, bringing it closer to our A$^3$COSMOS estimation. Nevertheless, both the acc\_time\_reducedx3 and acc\_time\_reducedx10 simulations overestimate the DMD at $z \lesssim 0.5$. 

$\;$
At higher redshifts, in the range probed by our data ($z > 1.5-2$), any of the reported models is able to reproduce the normalization and the shape of the observed DMD. In particular, the slope of the predicted DMDs is typically steeper than what is found observationally at $ z > 1.5 $. This lack of dust in high-$z$ galaxies in the models could be due to a lack of the most star-forming, IR-bright galaxies in the simulations, as found by \cite{gruppioni2015sam}, \cite{katsianis2017eagle}. From our analysis performed by modifying the model by \cite{parente2023dmdm_sim}, we argue that the currently implemented dust physics is not solely responsible for the observed discrepancies. Other sources of tension might arise from different physical processes incorporated in the models. This is further supported by the large variation among the other models, which are based on different galaxy evolution simulations. Building on this clue, we might speculate that one way of improving the models would then be to make a cross-comparison with a variety of different observables. For example, it would be interesting to investigate how the dust-richest high-$z$ galaxies appear in terms of SFR, size and morphology. Such a study would make it possible to improve the galaxy evolution model in several aspects, and hopefully achieve a better match with observations.

\par Since the A$^3$COSMOS DMD is derived from the integration of the observed DMF, it is also interesting to compare the latter for the aforementioned SAM-based experiments, as shown in Fig. \ref{fig:dmf_parente}. As previously mentioned, modifying the stellar production and switching off the SNe-driven grains destruction has a minimal impact on the predicted DMF. Conversely, enhancing the accretion efficiency brings the SAM predictions closer to the observed DMF, particularly at $z \gtrsim 2$ in the accx10 run. Among our experiments, this is the only way to generate objects with dust masses akin to those observed ($ \simeq 10^9 M_\odot$). We additionally note a discrepancy between the simulated DMFs and the extrapolated Schechter fit at $M_{\rm dust} \lesssim 10^8 M_\odot$, where direct observations are unavailable. 
\par We conclude this section with two caveats regarding our numerical experiments. The first concerns the underlying galaxy evolution framework of the model. Specifically, the SAM by \cite{parente2023dmdm_sim} has been noted to exhibit a lack of highly star-forming galaxies at $z \gtrsim 1$ (\textcolor{blue}{Traina, in prep.}). These galaxies also host substantial amounts of dust, the same dust missing in Fig. \ref{fig:dmf_parente}. Secondly, while increasing accretion alleviates discrepancies with observations, we caution that it may affect the relationship between the dust-gas ratio and ISM metallicity \citep{parente22}, a topic beyond the scope of this study.

\section{Summary and Conclusions}\label{summary}
In this paper, we investigated the properties of the dust mass budget in high-redshift galaxies. To this end, we studied a sample of 189 ALMA-selected star-forming galaxies in a wide redshift range ($0.5 < z < 6$), drawn from the \a3 database. By performing SED fitting analysis, we measured the dust content of each galaxy (i.e., dust mass and temperature) and we used these estimates to derive the DMF and the DMD. We summarize our results as follows:
\begin{itemize}
    \item The \a3 star-forming galaxies are dust-rich, with SED-based dust masses between $10^8$ and $10^9$ M$_{\odot}$ and the bulk of them showing a $T_{\rm D} \sim 30-35$K;
    
    \item RJ-based dust masses are in good agreement with the SED fit-based ones, when the assumed dust temperature is roughly consistent with those inferred from our SED fit ($\sim$35K), while the dust mass estimate with $T_{\rm D}=25$ are $\sim 54\%$ higher;
    
    
    \item The DMF inferred from the \md derived from the SED fit are in good agreement with those derived by \cite{pozzi2020dmd} in the mass and redshift range in common between these two studies;

    \item Combining the {\it Herschel} and ALMA DMFs, we inferred, for the first time, the evolution of the DMFs over a wide range of redshifts ($0.5 < z < 6$). The characteristic density ($\Phi^*$) and mass ($M_{D}^*$) of the DMFs are evolving with a decreasing ($\sim 2$ dex) and increasing ($\sim 1$dex) trend with redshift, respectively;
    
    \item {Integrating the DMFs, we found that the DMD evolves with a smoothly decreasing trend from $z \sim 0.75$ to $z \sim 5.25$, without showing a drastic drop towards higher redshifts.}

    \item None of the models available in literature is able to match both the shape and normalization of the observed DMD at $0.5<z<5$. Through dedicated numerical experiments, we find that the grains accretion in the ISM is the most effective dust-related process for increasing the dust content of galaxies.
    
\end{itemize}


\section*{Acknowledgements}
We would like to thank the anonymous referee for the positive and constructive report.
I.D. acknowledges funding by the European Union - NextGenerationEU, RRF M4C2 1.1, PRIN 2022JZJBHM: "AGN-sCAN: zooming-in on the AGN-galaxy connection since the cosmic noon" - CUP C53D23001120006
\bibliographystyle{aa}
\bibliography{1biblio}

\begin{appendix}

\end{appendix}

\end{document}